\documentclass[aps,twocolumn,nofootinbib,showpacs,final,balancelastpage,superscriptaddress]{revtex4}
\usepackage{graphicx}
\usepackage{epstopdf} 
\usepackage{tabularx}
\usepackage{amssymb}
\usepackage{amsmath}
\usepackage{amsbsy}
\usepackage{comment}
\usepackage{mathrsfs}
\usepackage{dsfont}
\usepackage{nicefrac}
\usepackage{wrapfig}
\usepackage{amsfonts}
\usepackage{url} 
\usepackage[nolist]{acronym}
\usepackage{nicefrac}
\usepackage{subfigure}
\usepackage{epsf,color,colordvi,pifont}
\usepackage{showkeys}
\DeclareFontFamily{OT1}{pzc}{}
\DeclareFontShape{OT1}{pzc}{m}{it}{<-> s * [1.10] pzcmi7t}{}
\DeclareMathAlphabet{\mathpzc}{OT1}{pzc}{m}{it}

\newcommand{\vecAA}{\pmb{\mathpzc A}}

\newcommand{\Nplat}{N_{\mathrm{plateau}}}

\begin{document}
\title{Simulating dynamically assisted  production of Dirac pairs\\ in gapped  graphene monolayers}
\author{I. \surname{Akal}}
\affiliation{II. Institute for Theoretical Physics, University Hamburg, Luruper Chaussee 149, 22761 Hamburg, Germany }
\author{R. \surname{Egger}}
\author{C. \surname{M\"{u}ller}}
\author{S. \surname{Villalba-Ch\'avez}}
\email{selym@tp1.uni-duesseldorf.de}
\affiliation{Institut f\"{u}r Theoretische Physik, Heinrich-Heine-Universit\"{a}t D\"{u}sseldorf,\\ Universit\"{a}tsstr.\,1, 40225 D\"{u}sseldorf, Germany}

\begin{abstract}
In a vicinity of the Fermi surface, graphene layers with  bandgaps allow for closely simulating the vacuum of quantum electrodynamics and, thus, its yet unverified 
strong-field phenomenology with accessible field strengths. This striking feature is exploited to investigate a plausible materialization of dynamically assisted 
pair production through the analog production of light but massive pairs of Dirac quasiparticles. The process is considered in a field configuration combining a weak 
high-frequency electric mode  and a strong low-frequency electric field oscillating in time. Its theoretical study is carried out  from a quantum kinetic approach, similar 
to the  one governing the spontaneous production of pairs  in QED. We show that the presence of the weak assisting  mode  can strongly increase the number of 
produced massive Dirac pairs as compared  with a setup driven by the strong field only. The efficiency of the process is contrasted, moreover, with the case of gapless 
graphene to highlight the role played by the quasiparticle mass.
\end{abstract}



\date{\today}

\maketitle

\section{Introduction}

Perceiving the vacuum as a region in which quantum fluctuations of all realizable fields in nature occur has always been accompanied by the intriguing possibility of 
their materialization into real particles at the expense of the vacuum instability. Among field theories describing the fundamental interactions, quantum electrodynamics (QED) 
provides the most accessible scenario in which this phenomenon could be realized. There, the spontaneous production of fermions is predicted to occur 
in form of  electron-positron pairs if an electric-like background [$E^2-B^2>0$] polarizes the quantum vacuum.\footnote{Here and henceforth the speed of light in 
vacuum and the absolute charge of an electron will be denoted by $c$ and $e$, respectively. Besides, $\hbar$ identifies the Planck constant.} The simplest configuration allowing for the pair production (PP) 
process -- commonly known as the Schwinger mechanism -- requires just a constant electric field [$B=0,\; E=\rm const$]. In such a situation, the associated PP rate $\mathcal{R}\sim \exp[-\pi m^2c^3/(e \hbar E)]$ 
manifests an essential singularity in the coupling strength $e$ indicating the nonperturbative nature of this striking strong field phenomenon \cite{Sauter:1931,Heisenberg:1935,Schwinger:1951nm}. 
That the Schwinger mechanism has so far eluded any detection agrees with the exponential suppression that $\mathcal{R}$ undergoes, because -- by now -- it is impossible to generate macroscopically extended  
electric fields with amplitude $E$ comparable to the critical scale in QED, $E_{\rm cr}=m^2c^3/(e\hbar)\sim 10^{16}\ \rm V/cm$. 

Due to continuous progress toward high-power lasers facilities, the possibility of observing the (dynamical) Schwinger mechanism in tightly focused pulses is currently receiving considerable  
attention. Envisaged multi-petawatt facilities such as the Extreme Light Infrastructure (ELI) \cite{ELI} and the Exawatt Center for Extreme Light Studies  (XCELS) \cite{xcels} 
are expected to achieve peak field strengths of the order of $\sim10^{-2}E_{\rm cr}$, which will bring the PP process closer to experimental accessibility. Motivated by these prospects, theoreticians
have studied special configurations of laser fields that might help to attain a detectable PP signal (see Refs.~\cite{Hebenstreit:2014lra,Bulanov2010} and references therein). In particular, various proposals  
have been put forward to mitigate the exponential suppression of the Schwinger pair creation  rate \cite{dgs2008,dgs2009,Orthaber2011,Grobe2012,Akal:2014eua,Otto:2014ssa,Linder:2015vta,Panferov:2015yda,torgrimsson2016,Akal:2017em}. 
Indeed, compelling theoretical evidences indicate a significant enhancement of the PP rate $\mathcal{R}\sim \exp[-\kappa E_{\rm cr}/E]$ with $0<\kappa\ll1$, when the strong field configuration is 
assisted by a fast-oscillating laser beam of weaker intensity.\footnote{Similar enhancement has been predicted to occur in production channels other than the one described 
so far, provided the assisting  high-frequency laser wave is present \cite{DiPiazza:2009py,Jansen2013,Augustin2014}.} In a physically intuitive picture, this enhancement in $\mathcal{R}$ is caused by the  absorption 
of a photon from the fast-oscillating field, this way reducing effectively the width of the barrier that an electron has to tunnel from negative to positive Dirac continuum. 

Although  promising, this so-called dynamically assisted Schwinger mechanism still awaits a  proof-of-principle test, and even with the envisaged fields in operation, it is  likely 
that its experimental verification remains a very challenging task. A complementary route for probing the vacuum instability assisted by an additional fast-oscillating wave is to 
look for a QED-like vacuum in which the mentioned process becomes manifest at much lower energies and field strengths. In this sense, graphene monolayers designed with a  bandgap 
$\Delta\varepsilon \lesssim 0.3\ \rm eV$ \cite{Substrate1,Substrate2,basov,Varykhalov} constitute an ideal scenario. Indeed, recent measurements of optical radiation emitted from regular 
graphene layers [$\Delta\varepsilon=0$]  irradiated by ultrashort terahertz pulses,  provide  strong evidence about  the  interband transition of electrons in the field of the THz pulse. The radiation 
is presumably caused by the recombinations of electron-hole pairs created via the Schwinger mechanism \cite{Oladyshkin}. See also Refs.~\cite{Lui,Tani}.

The suitability of bandgapped -- or semiconductor -- graphene layers  can be understood as a direct consequence of the quasiparticle mass $\mathpzc{m}=\Delta\varepsilon/(2\mathpzc{v}_f^2)\lesssim 10\ \mathrm{keV}/c^2$ 
[with $\mathpzc{v}_f\approx c/300$ the Fermi velocity] and the relativistic-like dispersion relation $\mathpzc{w}_{\pmb{p}}=(\pmb{p}^2\mathpzc{v}_f^2+\mathpzc{m}^2\mathpzc{v}_f^4)^{\nicefrac{1}{2}}$ 
that Dirac fermions exhibit when their energies are nearby the Fermi surface, i.e. close to any of the two inequivalent Dirac points $\pmb{K}_\pm$ in the reciprocal lattice \cite{Wallace}. 
Indeed, within the nearest-neighbours tight-binding model these two properties allow us to closely simulate the matter sector of QED, and thus, its quantum vacuum. Likewise, the 
electron-positron PP process finds an analogous phenomenon: the field-induced creation of quasiparticle-hole pairs from valence to conduction band. Recently, the similarity has been 
stressed further by noticing that the associated PP rate in the vicinity of $\pmb{K}_\pm$ closely resembles the exponential dependence $\mathcal{R}_{\mathpzc{g}}\sim \exp[-\pi \mathpzc{m}^2\mathpzc{v}_f^3/(e\hbar E)]$ 
found for the Schwinger mechanism \cite{Akal:2016stu}.  The feasibility for materializing relativistic-like tunneling in semiconductors has also been investigated in Ref.~\cite{Linder:2015fba}. 
As in QED,  critical field strengths can be associated with these materials. For instance, in graphene varieties with band gaps this one $E_{\mathpzc{g}}=\mathpzc{m}^2 \mathpzc{v}_f^3/(e \hbar)\lesssim 10^{5}\ \rm V/cm$ 
is set by the exponent of $\mathcal{R}_{\mathpzc{g}}$. However, in contrast to $E_{\mathrm{cr}}$, the latter can be easily overpassed with present-day laser technology, without damaging 
the material. This feature offers a genuine opportunity for a well-controllable materialization of both the Schwinger mechanism and its dynamically assisted version through the described
solid-state analog.

Here, we investigate the production of quasiparticle-hole pairs in an external field configuration involving a fast-oscillating electric mode of weak intensity superposed on a 
strong, low-frequency electric field oscillating in time. Our study is carried out from a quantum kinetic approach, similar to the one governing the spontaneous PP process in QED. 
Our outcomes reveal that the number of created massive quasiparticles in this assisted setup can be strongly enhanced by the presence of the high-frequency mode. In particular,  by 
exploiting the laser repetition rate  in our bifrequent setup, the number of pairs can become 
comparable to the single-shot outcome resulting from a scenario in which the high-frequency wave is not present and the Dirac fermions are massless instead. The mentioned aspects are exposed in two sections. In Sec.~\ref{sec:QKA}, we set the theoretical framework on  which our study relies. 
Likewise, some important features of the production of electron-hole pairs in bandgapped graphene are analyzed in connection with transport theory. The numerical results are presented 
in Sec.~\ref{sec:numerical}, where   in parallel, the role of the resonant effects arising in the single-particle distribution function is discussed. Also the density of pairs yielded in 
the assisted field configuration is shown in this section. Further comments and remarks are  given in the conclusions.

\section{Quantum kinetic approach \label{sec:QKA}}

Let us briefly precise some general features linked to the spontaneous production of electron-hole pairs in graphene layers. Previous studies on this phenomenon have been carried out 
\cite{Allor:2007ei,Mostepanenko,Avetissian,Fillion-Gourdeau:2015dga,Fillion-Gourdeau:2016izx} in connection with the absence of a gap $\Delta\varepsilon=0$ that the regular synthesization 
of graphene manifests \cite{GrExp1,GrExp2,GrExp3}. Such a characteristic constitutes, however, a drawback when thinking of materializing the dynamically-assisted PP process via the graphene 
analogue because the exponential suppression  of the corresponding rate is removed fully. As this behavior is essential for judging both the plausible PP enhancement as well as the nonperturbative 
character of this phenomenon, we will suppose that the charge carriers can acquire a tiny mass $\mathpzc{m}=\Delta\varepsilon/2\mathpzc{v}_f^2$ corresponding to a gap $\Delta\varepsilon\sim 0.1\;\rm eV$  
[$\lambda=2\pi/(\mathpzc{m}\mathpzc{v}_f)\sim 0.1\ \rm nm $], which can originate from the epitaxial growth of graphene on SiC substrate \cite{Substrate1,Substrate2}, for instance. We 
emphasize that, in addition to this techniques, there exist a variety of well-known forms to induce gaps in the band structure of graphene. They include elastic strain \cite{basov} 
or Rashba spin splittings on magnetic substrates \cite{Varykhalov}. 

From now on the external electric field $\pmb{E}(t)=-\partial \pmb{\mathpzc{A}}(t)/(c\partial t)$ [$\mathpzc{A}_0(t)=0$] is supposed spatially homogeneous  but time dependent with  
\begin{align}
\vecAA (t) = &-\mathscr{F}(\phi)\left[\frac{cE_s}{\Omega}\sin(\phi) +\frac{cE_w}{\omega}\sin(\eta \phi)\right]\pmb{n}.
\label{eqn:gauge-field}
\end{align} Here $\Omega$ and $E_s$ are the respective frequency and the electric field amplitude of the strong field, whereas  $\omega$ and $E_w$ correspond to the frequency and the  field 
strength of the perturbative fast-oscillating wave.  The parameters contained within the trigonometric functions are $\phi=\Omega t$ and $\eta=\omega/\Omega$. Besides, $\pmb{n}^T = (0,1,0)$ 
defines the polarization direction of the field which we assume -- hereafter -- embedded in the plane defined by the graphene sheet, i.e. at $z=0$. In Eq.~\eqref{eqn:gauge-field} the envelope 
function is chosen with $\sin^2$-ramping and de-ramping intervals, whereas a plateau region of constant field intensity is taken in between:
\begin{equation}
\begin{split}
\mathscr{F}(\phi)&=\sin^2\left(\phi/2\right) \left[\Theta(\pi-\phi)\Theta(\phi)+\Theta(2\pi N-\phi)\right.\\ &\times\left.\Theta(\phi-2\pi\mathpzc{K})\right]+\Theta(\phi-\pi)\Theta(2\pi\mathpzc{K}-\phi),
\end{split}
\label{eqn:envelope}
\end{equation} where $N =\Nplat + 1$,  $\mathpzc{K} = N - \frac{1}{2}$ and  $\Theta (x)$ denotes the unit step function: $\Theta (x)=1$ at $x\geqslant 0$, $\Theta (x)=0$ at $x<0$.  Here $\Nplat$ counts 
for the number of cycles within the region where $\mathscr{F}(\phi)=1$. Observe that Eq.~\eqref{eqn:gauge-field} and  \eqref{eqn:envelope} guarantee the  starting of the oscillating 
electric field  with zero-amplitude at $t=0$.  

There exist various ways of generating a field of this nature. A first option could be a head-on collision of two pairs of linearly polarized laser pulses sharing two frequencies, the same polarization 
directions and which propagate perpendicularly to the graphene sheet. A second option can be envisaged by combining the field generated by a capacitor with ac voltage and  the one generated by the collision 
of two counterpropagating  monochromatic laser waves which, as in the previous setup, impinge perpendicularly to the layer.  Observe that, to fit with our theoretical treatment, the field 
directions of both time dependent electric components have to coincide. The described scenarios do not provide a homogeneous electric field; those  resulting from the wave collisions depend on their direction 
of propagation. However, because of the chosen geometry, no essential effect due to the spatial inhomogeneity along the direction of propagation is expected. Clearly, the field generated by the  laser waves  
also has a transverse finite extension characterized by the waist size $w_0$, and so it is inhomogeneous in the  $(x,y)-$plane. We remark that no impact due to this transverse focusing is expected either as 
long as $w_0$ turns out to be much larger than the length of the graphene layer, which we take here of the order of $\ell\sim \mathcal{O}(10)\;\rm \mu m$.  Clearly, the counterpropagation of the 
electric-field-generating plane-waves could take place along the graphene sheet rather than perpendicular to it.  If this is the case, the homogeneous electric field approximation requires  the involved 
laser wavelengths $\lambda_{s,w}$ to  exceed  significantly the typical formation length of the PP process $\mathpzc{l}_{\mathrm{pair}}\sim \delta/\vert e E_{s}\vert$ with 
$\delta=\Delta\varepsilon-\omega$ \cite{DiPiazza:2009py,Jansen2013}.

From the theory prospect, quantum kinetic theory constitutes an appropriate approach for investigating the production of electron-hole pairs in graphene. In contrast to perturbative processes in unstable vacuum--described 
by Feynman diagrams \cite{Gitman,Fradkin}--the dynamical information of the pure PP process can be comprised in the single-particle distribution function  $W_{\mathpzc{g}}(\pmb{p},t)$ of electrons and holes to which the degrees of freedom in the external field are relaxed 
at asymptotically large times [$t\to\pm\infty$],  i.e., when  the electric field is switched off $\pmb{E}(\pm\infty)\to 0$.  Whenever the momentum of the massive quasiparticles $\pmb{p}$, relative to $\pmb{K}_\pm$  
satisfies the condition $\vert\pmb {p}\vert\ \ll \vert\pmb{K}_\pm\vert=\frac{4\pi}{3\sqrt{3}a_0}$ with  $a_0=0.142\ \rm nm$ \cite{basov}, the time evolution of this quantity is dictated by a quantum kinetic equation  
\cite{Akal:2016stu}:\footnote{In the following we set the Planck constant equal to unity, $\hbar=1$.}
\begin{eqnarray}\label{vlasovgraphene}
&&\dot{W}_{\mathpzc{g}}(\pmb{p},t)=Q(\pmb{p},t)\int_{-\infty}^t d\tilde{t}\; Q(\pmb{p},\tilde{t})
\left[\frac{1}{2}-W_{\mathpzc{g}}(\pmb{p},\tilde{t})\right]\nonumber\\&&\qquad\quad\qquad\times\cos\left[2\int_{\tilde{t}}^t dt^{\prime}\ \mathpzc{w}_{\pmb{p}}(t^{\prime})\right].
\end{eqnarray}Here,  the  vacuum initial condition $W_{\mathpzc{g}}(\pmb{p},-\infty)=0$ is  assumed. This formula is characterized by the function $Q(\pmb{p},t)\equiv e E(t)\mathpzc{v}_f \epsilon_\perp /\mathpzc{w}_{\pmb{p}}^2(t)$,  
which depends on the transverse energy of the Dirac fermions $\epsilon_\perp=\sqrt{\mathpzc{m}^2\mathpzc{v}_f^4+p_x^{\,2}\mathpzc{v}_f^2}$  and their respective  total energy  squared
$\mathpzc{w}^2_{\pmb{p}}(t)=\epsilon_\perp^2+[p_y-e\mathpzc{A}(t)/c]^2\mathpzc{v}_f^2$ of a Dirac quasiparticle.  As no dependence on temperature is manifested in Eq.~(\ref{vlasovgraphene}), 
any prediction resulting from it must be  understood within the zero temperature limit. We remark that, at asymptotically earlier times $t\to-\infty$ for which both $\mathpzc{A}(t),\;E(t)\to0$,  
the system is assumed to meet equilibrium conditions. In this limit  the total energy of a Dirac quasiparticle $\mathpzc{w}_{\pmb{p}}(t)\to\sqrt{\pmb{p}^2\mathpzc{v}_f^2+\mathpzc{m}^2\mathpzc{v}_f^2}$ 
has to be understood relative to the Fermi-level:  
\begin{equation}
\varepsilon_\mathrm{F}=\mathpzc{mv}_f^2\left[\sqrt{\frac{\mathpzc{N}_{\;\mathrm{in}}}{\mathpzc{N}_{\;0}}+1}-1\right].
\end{equation}Here $\mathpzc{N}_{\;\mathrm{in}}$ denotes the initial density of Dirac fermions,\footnote{As  occurs in standard gate-tunable setups,  $\mathpzc{N}_{\;\mathrm{in}}$ can be varied, 
and so  the Fermi level for the Dirac  fermions.} whereas the constant value  $\mathpzc{N}_{\;0}=g_sg_v \mathpzc{m}^2\mathpzc{v}_f^2/(4\pi)$  depends on the respective 
``spin'' ($g_s=2$) and valley ($g_v=2$)  degeneracy factors. For $\mathpzc{m}=11.7\; \mathrm{keV}/c^2$, corresponding to a bandgap  $\Delta\varepsilon=0.26\;\rm eV$,   
$\mathpzc{N}_{\;0}\approx 1.26\times 10^{12}\;\rm cm^{-2}$.  Observe that, in the limit $\mathpzc{N}_{\mathrm{in}}\gg \mathpzc{N}_{\;0}$, the Fermi energy reduces to the expression known 
for gapless graphene monolayers: $\varepsilon_{\mathrm{F}}\approx(\pi\mathpzc{v}_f^2\mathpzc{N}_{\;\mathrm{in}})^{\nicefrac{1}{2}}$.

Despite  its compactness, the consequences embedded in Eq.~(\ref{vlasovgraphene}) are often facilitated via  equivalent systems of ordinary differential equations. Here we utilize one combining two equations in which 
the Bogoliubov coefficients $f(\pmb{p},t)$ and $g(\pmb{p},t)$  appear explicitly \cite{Akal:2014eua,Akal:2016stu}:\footnote{Another representation involving three coupled ordinary differential equations can be found 
in the literature as well. See for instance Refs.~\cite{Hebenstreit:2010vz,Otto:2014ssa,Panferov:2015yda}.}
\begin{equation}\label{firstequa}
\begin{split}
&i\dot{f}(\pmb{p},t)=\mathpzc{a}_{\pmb{p}}(t)f(\pmb{p},t)+\mathpzc{b}_{\pmb{p}}(t)g(\pmb{p},t),\\ 
&i\dot{g}(\pmb{p},t)=\mathpzc{b}^*_{\pmb{p}}(t)f(\pmb{p},t)-\mathpzc{a}_{\pmb{p}}(t)g(\pmb{p},t).
\end{split}
\end{equation}In this context  the distribution function is given by $W_{\mathpzc{g}}(\pmb{p},t)=\vert f(\pmb{p},t)\vert^2$ and 
the initial conditions are chosen so that $f(\pmb{p},-\infty)=0$ and $g(\pmb{p},-\infty)=1$. The remaining  parameters contained in these formulas  are given by
\begin{eqnarray}\label{coefficient1}
\begin{array}{c}
\displaystyle\mathpzc{a}_{\pmb{p}}(t)=\mathpzc{w}_{\pmb{p}}(t)+\frac{eE(t)p_x\mathpzc{v}_f^2}{2\mathpzc{w}_{\pmb{p}}(t)(\mathpzc{w}_{\pmb{p}}(t)+\mathpzc{m}\mathpzc{v}_f^2)},\\ \\
\displaystyle\mathpzc{b}_{\pmb{p}}(t)=\frac{1}{2}\frac{eE(t)\epsilon_\perp}{\mathpzc{w}_{\pmb{p}}^2(t)}\exp\left[-i\tan^{-1}\left(\frac{p_x q_\parallel\mathpzc{v}_f^2}{\epsilon_\perp^2+\mathpzc{m}\mathpzc{v}_f^2\mathpzc{w}_{\pmb{p}}(t)}\right)\right],\\
\end{array}
\label{coefficient2}
\end{eqnarray}where the kinetic momentum along the external field direction is $q_\parallel=p_y-e\mathpzc{A}(t)/c$. 

Further aspects deserve to be mentioned. It is worth remarking that Eq.~(\ref{vlasovgraphene})  -- or alternatively Eq.~(\ref{firstequa}) --  strongly relies on the electron-hole symmetry. Hence, its outcomes should 
provide good estimates within the nearest-neighbours tight-binding model. However, if the next-to-nearest-neighbours interactions are taken into account this invariance does not hold anymore \cite{Wallace}  
and corrections due to this fact  have to be included. We note besides that the treatment based on these formulae ignores the effect caused by both the collisions between the created Dirac-quasiparticles and their inherent 
radiation fields. Similar to QED \cite{Tanji:2008ku,Bloch:1999eu,Vinnik:2001qd}, both phenomena are expected to become relevant as the field strength  is stronger than the 
critical field scale in bandgapped graphene $E\gtrsim E_{\mathpzc{g}}$ with  $E_{\mathpzc{g}}= \mathpzc{m}^2\mathpzc{v}_f^3/e\sim 10^5\ \rm V/cm$. However, for  assessing the  plausible materialization of the assisted Schwinger 
mechanism in band-gapped graphene,  field strengths $E_{s,w}<E_\mathpzc{g}$ will  be utilized.

Noteworthy, as the production of electron-hole pairs in graphene is governed by a transport equation [see Eq.~(\ref{vlasovgraphene})] which resembles the one in QED, various existing insights  
linked to $W_{\mathrm{QED}}(\pmb{p},t)$ can be extrapolated to elucidate the behavior of $W_{\mathpzc{g}}(\pmb{p},t)$. Thus, we expect that $W_{\mathpzc{g}}(\pmb{p},t)$ mimics 
the resonances spectrum  \cite{Akal:2014eua,Otto:2014ssa,Panferov:2015yda}  associated with the absorption of quanta from the field [see Eq.~(\ref{eqn:gauge-field})]. This  phenomenon takes place as the resonant 
condition 
\begin{equation}\label{rensonantcondition}
 2\bar{\varepsilon}_{\pmb{p}}\approx n_s\Omega+n_w\omega
\end{equation}holds. In this relation,  $n_s$ ($n_w$) refers to the number of absorbed  quanta from the strong (weak) wave,  whereas $\bar{\varepsilon}_{\pmb{p}}=\frac{1}{\tau} \int_{0}^{\tau}dt\;\mathpzc{w}_{\pmb{p}}(t)$ 
denotes the quasienergy of the produced particles, i.e. the energy averaged over the total pulse length  $\tau$. The behavior of the distribution function $W_{\mathrm{QED}}(\pmb{p},t)$ near a resonance 
characterized by $n_s$ and $n_w$ is  known.  Assuming that the external  field exists for a time duration  $\tau$ we have
\begin{equation}
\begin{array}{c}\displaystyle
W_{\mathpzc{g}}^{(n_s,n_w)}(\pmb{p},\tau)\approx\frac{1}{4}\mathcal{A}_{\pmb{p}}^{(n_s,n_w)}\frac{\sin^2\left(x_{\pmb{p}}\right)}{x_{\pmb{p}}^2},\\ \displaystyle
\mathcal{A}_{\pmb{p}}^{(n_s,n_w)}=\vert\Lambda^{(n_s,n_w)}_{\pmb{p}}\vert^2 \tau^2,\quad x_{\pmb{p}}=\Omega^{(\mathrm{Rabi})}_{\pmb{p}}\tau.
\end{array}
\label{resonantdistributionfunction}
\end{equation} Here $\Lambda^{(n_s,n_w)}_{\pmb{p}}$ is a complex  time-independent coefficient whose explicit expression  is  not necessary to understand what follows. 
In this formula  the  Rabilike frequency of the vacuum $\Omega_{\pmb{p}}^{(\mathrm{Rabi})}=\frac{1}{2}\left[\vert\Lambda_{\pmb{p}}^{(n_s,n_w)}\vert^2+(\Delta_{\pmb{p}}^{(n_s,n_w)})^2\right]^{\nicefrac{1}{2}}$  
with   $\Delta_{\pmb{p}}^{(n_s,n_w)}\equiv2\bar{\varepsilon}_{\pmb{p}}-n_s\Omega-n_w\omega$ being  the detuning parameter. 
We should, however, emphasize that the above resonant approximation is valid if the Rabilike frequency  is slow in comparison with the laser one $\Omega^{(\mathrm{Rabi})}_{\pmb{p}}\ll\omega,\;\Omega$ 
\cite{mocken2010}. Observe that for a certain parameter combination  ($\Omega,\;\omega\; n_s\; n_w$), there might be multiple choices of momenta $\{\pmb{p}_0\}$ with $\pmb{p}_0=(p_{x0},p_{y0})$ 
which satisfy the resonant condition in Eq.~(\ref{rensonantcondition}). For all these points the detuning parameter vanishes [$\Delta^{(n_s,n_w)}_{\pmb{p}_0}\approx0$], the Rabilike frequency reduces to 
$\Omega^{(\mathrm{Rabi})}_{\pmb{p}_0}\approx \frac{1}{2} \vert\Lambda^{(n_s,n_w)}_{\pmb{p}_0}\vert$ and the single-particle distribution function approaches  
$W_{\mathpzc{g}}^{(n_s,n_w)}(\pmb{p}_0,\tau)\approx \sin^2\left[\Omega^{(\mathrm{Rabi})}_{\pmb{p}_0}\tau\right]$.
 
\section{Enhanced  production of  electron-hole pairs in graphene\label{sec:numerical}}

We wish to illustrate the phenomenological consequences of the external field [see  Eq.~(\ref{eqn:gauge-field})] on the PP process of massive charge carriers. In line with Ref.~\cite{Akal:2016stu} we chose 
$\Delta\varepsilon=0.26\;\rm eV$ as our standard band-gap value, which corresponds to  $\mathpzc{m}= 11.7\ \mathrm{keV}/c^2$. A suitable realization of the assisted scenario  requires, in first instance, a  
frequency  for the perturbative wave carrying a substantial fraction of band gap energy [$\omega\lesssim \Delta\varepsilon$]. The strong field frequency $\Omega$, on the other hand,  should remain smaller 
than the difference between  $\Delta\varepsilon$ and $\omega$, i.e., $\Omega<\Delta\varepsilon-\omega$. Because of this, we will take  $\Omega=0.026\ \rm eV$  and $\omega=0.206\ \rm eV$, so that $n_s\geqslant3$ 
is required for any field strength. The plateau region will comprise $\Nplat=26$ cycles corresponding to a total pulse length of  $\tau=2\pi N/\Omega\approx 4.2\ \rm ps$. Here, the strong field strength is 
taken $E_s= 4.9\times 10^4\ \rm  V/cm$ which corresponds to a laser intensity $I_s=cE_s^2= 6.4\times 10^6\ \rm W/cm^2$. The intensity of the perturbative laser wave is set to $I_w= 9.6 \times 10^4\ \rm W/cm^2$, 
leading to a field strength $E_w= 0.6\times 10^4\ \rm V/cm$.  Observe that, in the current  scenario the critical field is $E_\mathpzc{g}=\mathpzc{m}^2\mathpzc{v}_f^3/e\approx 2.6 \times 10^5\ \rm V/cm$ and, correspondingly, a critical scale for the laser intensity 
$I_\mathpzc{g}=cE_\mathpzc{g}^2\approx 1.8 \times 10^8\ \rm W/cm^2$ can be set.  Clearly,  the parameters above can be combined separately  to  define the  laser intensity parameter of the strong field 
$\xi_s=eE_s/(\mathpzc{mv}_f\Omega)\approx0.94$ and the perturbative fast-oscillating field $\xi_w=eE_w/(\mathpzc{mv}_f\omega)\approx0.014$. It is worth emphasizing that none of the chosen intensities for the waves overpasses the  critical scale  $I_s,\; I_w\ll I_{\mathpzc{g}}$. 
Indeed, all of them can be attained comfortably with terahertz laser systems with picosecond duration \cite{tera1,tera2}.  Noteworthy, both the wavelengths associated with the strong field $\lambda_s\approx47.4\;\rm \mu m$ 
and the highly oscillating pulse $\lambda_w\approx 5.9\;\rm \mu m$ exceed the characteristic formation lengths $\mathpzc{l}_{\mathrm{pair}}\sim  10^{-2}\;\mu \rm m$ [read discussion in the paragraph  
above the one containing   Eq.~(\ref{vlasovgraphene})].

\begin{figure}
\includegraphics[width=0.47\textwidth]{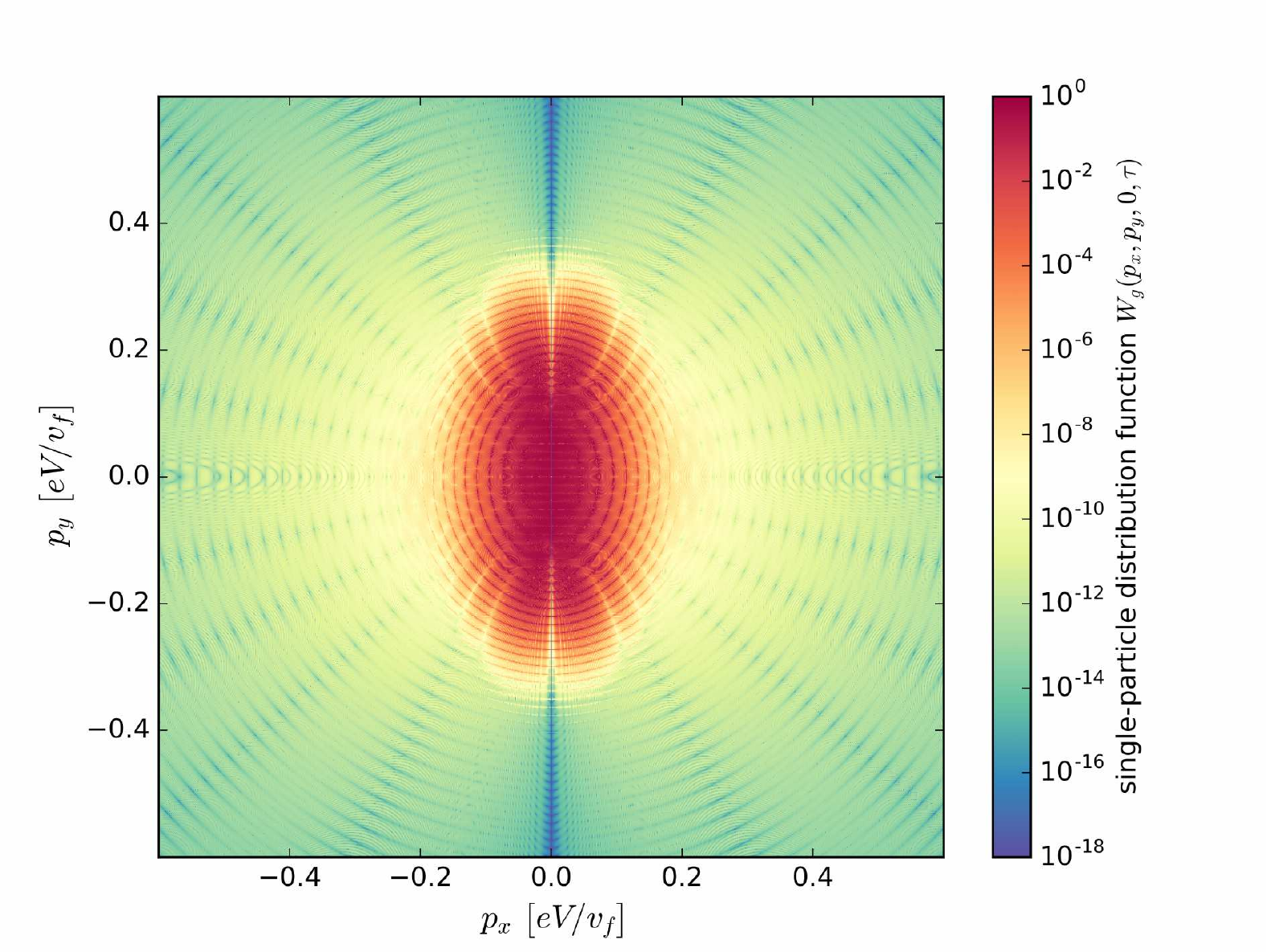}
\includegraphics[width=0.47\textwidth]{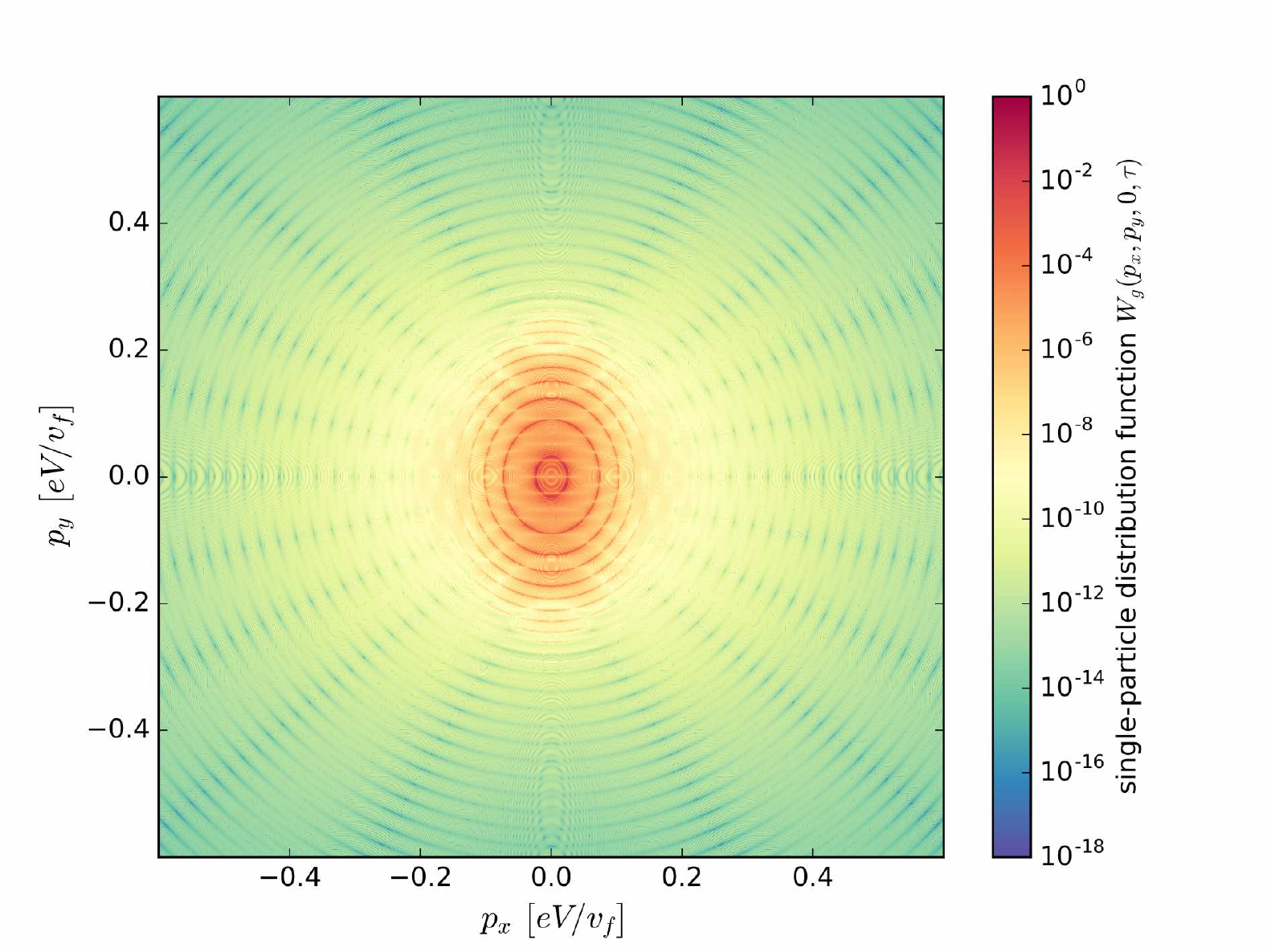}
\includegraphics[width=0.47\textwidth]{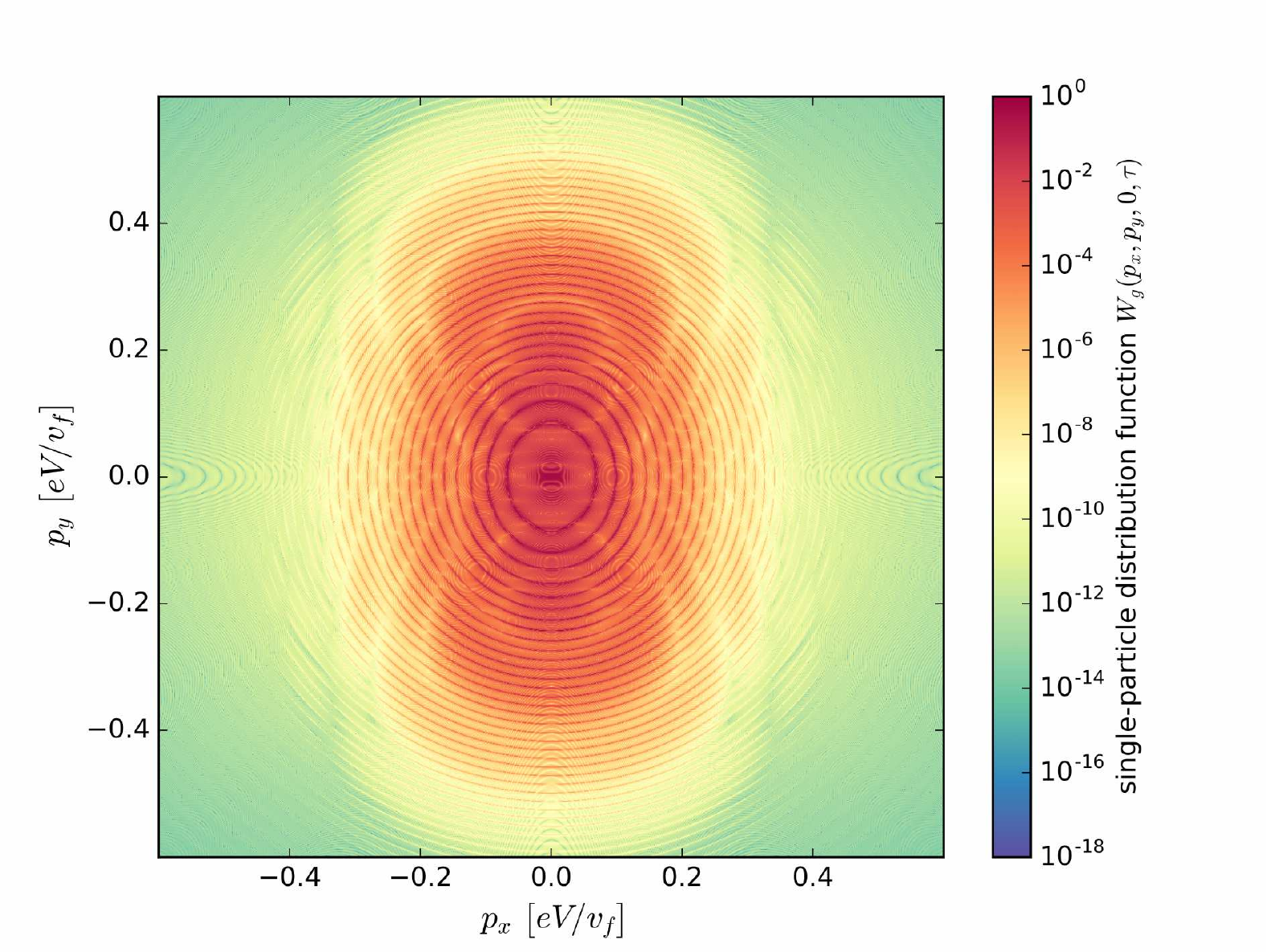}
\caption{Logarithmic plot of  the  single-particle distribution functions $W_{\mathpzc{g}}(\pmb{p},\tau)$. The result shown in the upper panel corresponds to a model involving massless charge carriers.  
Conversely, the middle and lower panels display outcomes obtained by considering massive Dirac quasiparticles. The distributions  in the upper and middle panels follow from the strong field only,  
i.e. $E_w = 0$, while the lower one takes into account  that both the strong and the fast-oscillating waves are present. The following  benchmark parameters have been used $\Delta \varepsilon=0.26\;\rm eV$, 
$E_s= 4.9\times 10^4\ \rm V/cm$, 
$E_w= 0.6\times 10^4\ \rm V/cm$,  $\Nplat=26$ cycles ($\tau\approx 4.2\ \rm ps$), $\Omega=0.026\ \rm eV$  and $\omega=0.206\ \rm eV$.}
\label{fig:1}
\end{figure}

\begin{figure*}
\includegraphics[width=.48\textwidth]{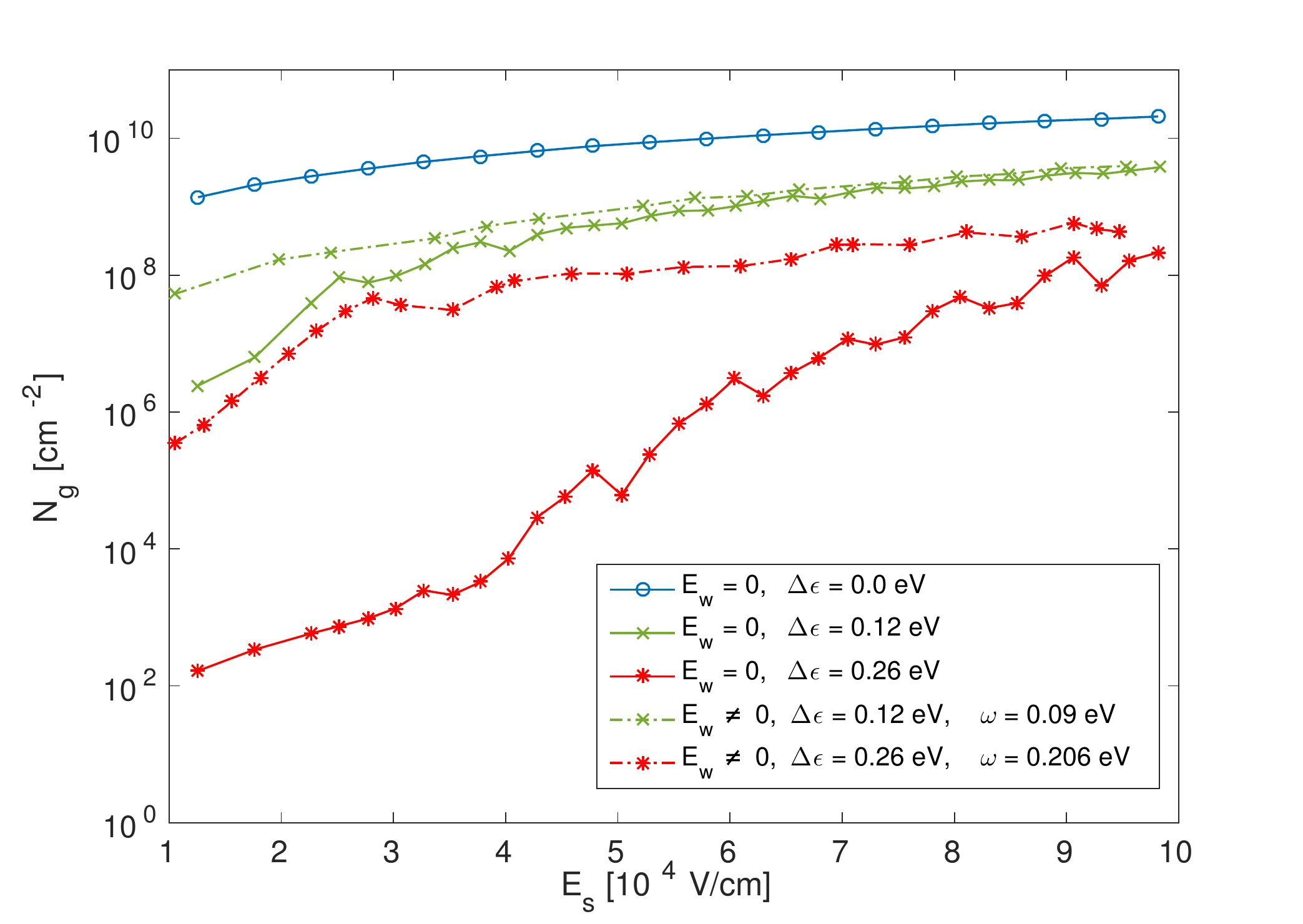}
\includegraphics[width=.49\textwidth]{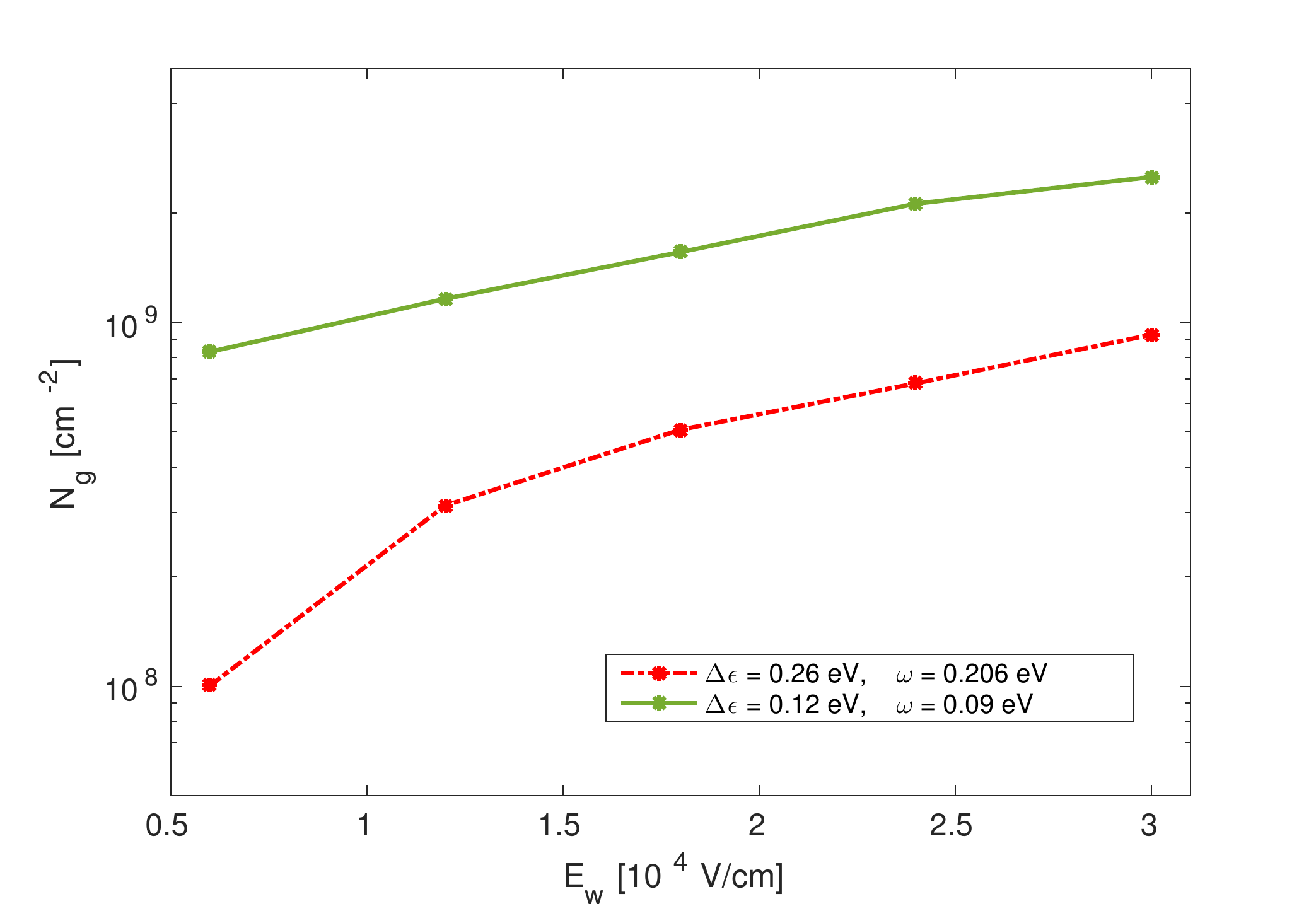}
\caption{Density of produced electron-hole pairs as a function of $E_s$[left panel]  and  $E_w$[right panel]. The blue curve in the left panel reveals the trend obtained for the massless model when the weak field is off.  
For the same field setting,  the lower curves in green and red -- both with more pronounced slopes as compared with the remaining curves -- result from  considering massive Dirac quasiparticles with bandgaps 
$\Delta\varepsilon=0.12\ \rm eV$ and $\Delta\varepsilon=0.26\ \rm eV$, respectively. The corresponding outcomes resulting from the configuration assisted by the perturbative fast-oscillating wave share the previous 
color scheme but the curves are those  containing  asterisks. Both  assisted curves have been  obtained by setting $E_w=0.6\times 10^4\;\rm V/cm$. The results exhibited in both panels were obtained by setting 
the  pulse length  $\Nplat=26\; \rm cycles$  ($\tau\approx4.2\; \rm ps$) and the strong field frequency  $\Omega=0.026\ \rm eV$. Besides,  for  $\Delta\varepsilon=0.26\ \rm eV$  the frequency of the perturbative field was chosen 
$\omega=0.206\ \rm eV$, whereas for  $\Delta\varepsilon=0.12\ \rm eV$  we have utilized  $\omega=0.09\ \rm eV$.  The curves in the right panel have been obtained by considering the two previous gaps: green 
for $\Delta\varepsilon=0.12\;\rm eV$ and red for  $\Delta\varepsilon=0.26\;\rm eV$. There the strong field was taken as $E_s= 4.9 \times 10^4\ \rm V/cm$ with  $\Omega=0.026\ \rm eV$. }
\label{fig:2}
\end{figure*}

Now, we solve the system of differential equations (\ref{firstequa}) by varying the momentum components within the interval $- 0.5\  \mathrm{eV}\leqslant p_{x,y}\mathpzc{v}_f\leqslant 0.5\  \mathrm{eV}$. The 
results of this analysis are displayed in Fig.~\ref{fig:1}  in a density color scheme which corresponds to $\log_{10}[W_\mathpzc{g}(\pmb{p},\tau)]$.  While the upper panel shows the  outcomes associated with the 
massless model, the consequences associated with the mass $\mathpzc{m}=11. 7\ \mathrm{keV}/c^2$ can be evaluated by comparing with the middle panel. It is worth remarking that both pictures were 
generated by considering the effects due to the strong field only. Their main features -- including the occurrence of a narrow  vertical blue sector in the massless case -- have been studied previously. For details, 
we refer the reader to  Ref.~\cite{Akal:2016stu}.  We note that these panels have been added here to facilitate -- via comparisons -- the understanding of the main outcome resulting from a scenario in which massive 
Dirac quasiparticles are produced by the combined effect of both the strong and the perturbative fast-oscillating wave [lower panel]. Observe that the three spectral densities are sharply anisotropic with respect to 
the momentum components, and manifest  gradual decrements as both $p_x$ and $p_y$  increase. The stretching along $p_y$ is understood as a result of the minimal coupling between the external field and the 
quasiparticle via the  kinetic momentum $p_y-e\mathpzc{A}(t)/c$, which enters into the quasienergy [see below Eq.~(\ref{rensonantcondition})]. Hence, the creation of  quasiparticle-hole pairs with rather large 
longitudinal momentum is more likely to take place. 

All  panels in Fig.~\ref{fig:1} are characterized by ringlike structures which differ between each other. Each ring is understood as an isocontour of fixed quasienergy $\bar{\varepsilon}_{\pmb{p}}$ satisfying the resonance
condition in Eq.~(\ref{rensonantcondition}). The regions encompassed between two neighboring rings are characterized by less intense colors, this way indicating the formation of intermediate 
valleys. Observe that the outcome shown in the middle panel, i.e. for massive carriers  with the fast-oscillating field turned off, looks rather different as compared with the model of massless Dirac quasiparticles 
[upper]. Clearly the number of resonances in the former is considerably smaller than in the latter case where the red-colored area  is much more pronounced and compact. In view of this behavior, we can infer that the 
volume below the surface $W_{\mathpzc{g}}(p_x,p_y)$ will be substantially larger in the massless model [upper panel] than in the case where the Dirac quasiparticles are massive and only the strong field is on [middle panel]. 
As these volumes result from the  integration  over the momentum components, actually they  will determine the number of yielded pairs  per unit of area  
\begin{equation}
\begin{split}
\mathpzc{N}_{\;\mathpzc{g}}=&\lim_{t\to\infty}g_sg_v\int \frac{d^2p}{(2\pi)^2} \, W_{\mathpzc{g}}(\pmb{p},t).
\end{split}
\label{eqn:tot-no}
\end{equation}  
We remark that  the factor  $g_s=2$ ($g_v=2$) accounts for  the spin (valley) degeneracy. Formally, the integration has to be performed over all the Fourier space. However, the outcomes given 
in Fig.~\ref{fig:1} provide evidences that the contribution of $W_{\;\mathpzc{g}}(\pmb{p},\tau)$ is  almost insignificant as long as  $p_x,\;p_y>10\; \mathrm{eV}/\mathpzc{v}_f$. This fact is consistent with the limitation 
indicated  above Eq.~(\ref{vlasovgraphene}). Hence, to compute Eq.~(\ref{eqn:tot-no}) numerically we will restrict the integration domain to an  area  $A$  embedded within a circle with radius 
$\vert\pmb{p}\vert\ll10 \ \mathrm{eV}/\mathpzc{v}_f$.

The situation changes when massive Dirac quasiparticles experience both the strong and the perturbative fast-oscillating wave simultaneously  [lower panel]. Owing to the latter, the number of resonances increases and the red area 
enlarges considerably as compared with the panel in the middle. This fact makes clear that the corresponding volume below the surface $W_{\;\mathpzc{g}}(p_x,p_y)$, i.e. the density of produced electron-hole pairs, 
increases in comparison to the situation where $\mathpzc{m}= 11.7\ \mathrm{keV}/c^2$ with the perturbative wave being switched off. Certainly, the red area in the lower panel turns out to be less intense than in the upper one 
[$\mathpzc{m}=0$]. However, in the former case, the red portion covers regions in the $(p_x,p_y)$-plane that the upper panel does not. Consequently, the volume enclosed by $W_{\;\mathpzc{g}}(p_x,p_y)$ in the 
assisted setting could reach values comparable to the one associated with the scenario involving massless carriers and, so for the density of created pairs [see Eq.~(\ref{eqn:tot-no})]. 

In order to verify this enhancement, we compute numerically the dependence of $\mathpzc{N}_{\;\mathpzc{g}}$ [see Eq.~(\ref{eqn:tot-no})] with respect to  the strong electric field strength $E_s$. The results of 
this assessment are summarized in the left panel of  Fig.~\ref{fig:2}. The outcome linked to the massless model is shown in blue, whereas those resulting from scenarios driven by massive Dirac quasiparticles characterized 
by a bandgap $\Delta\varepsilon=0.26\ \rm eV$  are depicted in red. For comparison, the results linked to a reduced gap  $\Delta\varepsilon=0.12\ \rm eV$ have  been included [green curves with an adjusted fast-oscillating frequency $\omega=0.09\;\rm eV$]. We note that the latter 
corresponds to a mass $\mathpzc{m}=5.4\;\mathrm{keV}/c^2$  and  a critical field strength $E_{\mathpzc{g}}\approx5.5\times 10^4\ \rm V/cm$ which is an order of magnitude smaller than the one associated with the 
energy gap $\Delta\varepsilon=0.26\ \rm eV$, i.e. $E_\mathpzc{g}\approx 2.6 \times 10^5\ \rm V/cm$. The right panel in Fig.~\ref{fig:2} shows the dependence of $\mathpzc{N}_{\;\mathpzc{g}}$  on the weak field 
strength $E_w$. Each curve corresponds to  one of the nontrivial bandgaps: green for $\Delta\varepsilon=0.12\;\rm eV$ and red for  $\Delta\varepsilon=0.26\;\rm eV$. 
Both have been obtained by setting the strong field to $E_s= 4.9 \times 10^4\ \rm V/cm$. As the  previous strength   is comparable to the critical field linked to the gap 
$\Delta\varepsilon=0.12\;\rm eV$, the corresponding results have to be considered as estimates.   Referring to the red curve, we see that the density of pairs  grows from left to right by a factor of $\approx 20$ when 
the amplitude of the weak field is increased by a factor of 8, implying an average growth which is faster than linear. For comparison we note that, for PP by a bichromatic electric field in a pure QED context 
\cite{Akal:2014eua}, a linear dependence on $E_w$ (respectively $\xi_w$) has been obtained to a good approximation. This simple scaling behavior may be attributed to the range of field parameters considered there.

Now, throughout the whole range of $E_s$ [left panel] the number of  pairs yielded  from  the  assisted scenario -- green and red curves marked with asterisks -- is  substantially larger than those arising when the perturbative 
fast-oscillating field is off, i.e. green and red curves marked with crosses. We note that, in a scenario with an energy gap  $\Delta\varepsilon=0.26\ \rm eV$ and  $E_s \approx2.7\times 10^4\;\rm V/cm$, the result of 
the assisted setup exceeds the outcome driven by a pure strong field setup by  four orders of magnitude. For comparison we note that the presence of the assisting field [$E_w=0.6\times 10^4\; \rm V/cm$, $\omega=0.206\; \rm eV$] enhances the pair  yield in the massless model 
by only a factor $2$, approximately. As -- in average -- the slopes of those curves linked to a pure strong field scenario [$\mathpzc{m}\neq0$] are larger than the  corresponding one in the assisted setting, the enhancement effect 
due to the weak field diminishes as $E_s$ grows [see also Refs.~\cite{dgs2008,DiPiazza:2009py,Jansen2013,Augustin2014}]. This behavior can be  expected because -- in an assisted scenario -- there 
are two paths for increasing the number of produced pairs. Namely, either by growing the strong field strength or through the absorption of energetic quanta, this way promoting the emergence of new resonances. While 
the former becomes more relevant as $E_s$ grows, the latter dominates as the opposite situation takes place.

Noteworthy, those curves linked to  scenarios in which the production of massive electron-hole pairs is driven by the strong field only, manifest  pronounced steep fallings which are caused by  
the so-called channel-closing effect [see, e.g., \cite{HReiss,Kohlfurst:2013ura,Akal:2014eua,Akal:2016stu}]. Also the results associated with the assisted setups reveal these drop-offs, although 
much less noticeable due to the large scale covered by the vertical axis. We recall that the channel-closing phenomenon can be attributed to the extinction of a resonance at those  field strengths 
for which $W_{\;\mathpzc{g}}(\pmb{p},\tau)$ is locally minimized.  Indeed, in the absence of a perturbative fast-oscillating wave, the number of produced pairs can increase through the growing 
of the strong field strength $E_s$. When this happens, the quasienergies that characterize the resonances  [see below Eq.~(\ref{rensonantcondition})] change too.  Most of theses resonances -- featured by a specific 
$n_s$ -- survive during the electric field change because -- when integrating over $\pmb{p}$ -- a readjustment in the corresponding momentum occurs in such a way that the relations $2\varepsilon_{\pmb{p}}=n_s\Omega$ still 
hold. However, throughout the change, the minimal energy to be absorbed from the field -- $\varepsilon_{\pmb{p}=0}$ -- might die out, this way compensating the enhancement induced by the strong field increment. If this 
occurs, the volume below the surface $W_{\;\mathpzc{g}}(p_x,p_y)$ diminishes and a drop-off emerges in the plane $(\mathpzc{N}_{\;\mathpzc{g}},E_s)$. That this behavior does not manifest in the assisted curves 
may be  then understood as a direct consequence  of the additional production channel that the assisted scenario provides. Within the context of the resonant condition [see Eq.~(\ref{rensonantcondition})], this manifests via 
the  term which contains  the number of quanta  $n_w$ absorbed from the weak field. 

Let us finally estimate the number of electron-hole pairs produced by  a single shot in  a sheet of graphene with  a characteristic length $\ell\sim 10\;\mu\rm m$. Consider first the nonassisted scenario with 
$E_s\approx 5.6\times 10^{4}\; \rm V/cm$ and $\mathpzc{m}=11.7\;\mathrm{keV}/c^2$, for which $\mathpzc{N}_{\;\mathpzc{g}}\sim 10^6\;\rm cm^{-2}$. Under such circumstances the number of pairs created from 
the vacuum in a single shot is  $\mathpzc{N}_{\;\mathpzc{g}}\ell^2\sim1$  approximately. However, if the previous setup is assisted by a perturbative fast-oscillating wave  with  $\omega=0.206\ \rm eV$ and 
$E_w\approx 3 \times 10^4\ \rm V/cm$, the density of pairs reaches va\-lues of the order of $\mathpzc{N}_{\;\mathpzc{g}}\sim 10^8\;\rm cm^{-2}$. As a consequence the number of electron-hole pairs produced by  
a single shot would be of the order of $N_{\mathrm{pairs}}\approx\mathpzc{N}_{\;\mathpzc{g}}\ell^2\sim10^2$, i.e. two orders of magnitude larger than in the nonassisted configuration. The layer could accumulate 
a much larger number of electron-hole pairs by taking advantage of  the  laser repetition rate. For instance,  values of the order of $N_{\mathrm{pairs}}\sim10^{6}$ could be reached after $n_{\mathrm{shots}}\sim 10^4$ laser shots. 
For a repetition rate of $\nu\sim\mathcal{O}(10-100)\;\rm Hz$, this  would require to irradiate the sample  during  $\mathcal{O}(2-17)\;\rm min$, approximately. After this period the density of produced electron-hole 
pairs reaches values of the order of $\mathpzc{N}_{\;f}=n_{\mathrm{shots}}\mathpzc{N}_{\;\mathpzc{g}}\sim 10^{12}\;\rm cm^{-2}$ which exceeds by $2$ orders of magnitude the density obtained from a single shot in 
a model with no gap [see  blue curve, left panel in Fig.~\ref{fig:2}].

We note that  for a final density $\mathpzc{N}_{\;f}\sim 10^{12}\;\rm cm^{-2}$  the electrostatic interactions  becomes relevant \cite{Lewkovicz,Ahiezer}. This way 
facilitating the attraction between carriers with different charges and, thus, the formation of excitons i.e. the  analog of the positronium in a pure QED context. As in the latter scenario, the formed excitons are 
expected to be unstable bound states which decay into photons. In first instance,  the detection of these photons would constitute a  direct signature of a prior realization of the assisted pair creation process. A 
study of the annihilation channels is beyond the scope of the present investigation, but they  will be analyzed in a forthcoming publication.

\section{Conclusion}

Matter-antimatter pair creation via the Schwinger mechanism is one of the most striking nonperturbative phenomena in quantum field theory. Since it has eluded any experimental verification, so far,
catalyzing mechanisms in nonstatic macroscopic gauge fields prove very promising in order to circumvent the  dramatic suppression of the creation rate. Alternatively, one may also look for analog 
realizations in appropriate low energy condensed matter systems like  graphene. In contrast  to earlier studies, we have focused on scenarios where a tiny bandgap is 
present in the vicinity of its Dirac points,  giving rise to the characteristic  tunneling exponential  in the Schwinger pair creation rate.  This feature has allowed us to verify theoretically 
that such semiconducting graphene varieties are rather suitable not only to probe the Schwinger effect but also the proposed mechanism of dynamical assistance.

Employing a quantum kinetic approach with reduced effective dimensionality, we first have computed the corresponding single-particle distribution function for experimentally motivated field parameters. 
In doing so, we have verified that the presence of an assisting weak field can yield a notably enhancement, approaching values obtained for ordinary massless graphene via the standard PP mechanism.
For all configurations a characteristic resonant behavior, reflected in form of isocontours of fixed quasienergy,  has been identified in the momentum plane.  Based on the obtained distribution functions, 
we then have computed the total number of created quasiparticle-hole pairs. We have shown that the latter enabled via the assisted mechanism turned out to be several orders of magnitude larger than 
in the nonassisted process. We have argued that, with a suitable laser repetition rate, the density of produced electron-hole pairs  increases by several orders  of magnitude, allowing to reach the level 
from  which the emission of photons -- as a result of the recombination process -- becomes a clear signature  of the assisted Schwinger  mechanisms.


\section*{Acknowledgments}

The authors thank  A.~Golub for useful discussions. S. Villalba-Ch\'avez and C.~M\"{u}ller  gratefully acknowledge the funding by the German Research 
Foundation (DFG) under Grant No. MU 3149/2-1.  I.~Akal acknowledges the support of the Colloborative Research Center SFB 676 of the DFG.


\end{document}